\documentclass{aastex63}

\shorttitle{}
\shortauthors{Kou et al.}

\begin{document}

\title{What determine Solar Flares Producing Interplanetary Type III Radio Bursts?}

\correspondingauthor{Xin Cheng}\email{xincheng@nju.edu.cn}

\author{Y. K. Kou}
\affiliation{School of Astronomy and Space Science, Nanjing University, Nanjing 210023, China\\}
\affil{Key Laboratory of Modern Astronomy and Astrophysics (Nanjing University), Ministry of Education, Nanjing 210093, China\\}

\author{Z. C. Jing}
\affiliation{School of Astronomy and Space Science, Nanjing University, Nanjing 210023, China\\}
\affil{Key Laboratory of Modern Astronomy and Astrophysics (Nanjing University), Ministry of Education, Nanjing 210093, China\\}

\author[0000-0003-2837-7136]{X. Cheng}
\affil{School of Astronomy and Space Science, Nanjing University, Nanjing 210023, China\\}
\affil{Key Laboratory of Modern Astronomy and Astrophysics (Nanjing University), Ministry of Education, Nanjing 210093, China\\}
\affil{Max Planck Institute for Solar System Research, Gottingen, 37077, Germany\\}

\author{W. Q. Pan}
\affiliation{School of Astronomy and Space Science, Nanjing University, Nanjing 210023, China\\}
\affil{Key Laboratory of Modern Astronomy and Astrophysics (Nanjing University), Ministry of Education, Nanjing 210093, China\\}

\author{Y. Liu}
\affiliation{School of Astronomy and Space Science, Nanjing University, Nanjing 210023, China\\}
\affil{Key Laboratory of Modern Astronomy and Astrophysics (Nanjing University), Ministry of Education, Nanjing 210093, China\\}

\author{C. Li}
\affiliation{School of Astronomy and Space Science, Nanjing University, Nanjing 210023, China\\}
\affil{Key Laboratory of Modern Astronomy and Astrophysics (Nanjing University), Ministry of Education, Nanjing 210093, China\\}

\author[0000-0002-4978-4972]{M. D. Ding}
\affiliation{School of Astronomy and Space Science, Nanjing University, Nanjing 210023, China\\}
\affil{Key Laboratory of Modern Astronomy and Astrophysics (Nanjing University), Ministry of Education, Nanjing 210093, China\\}

\begin{abstract}
Energetic electrons accelerated by solar flares often give rise to type III radio bursts at a broad waveband and even interplanetary type III bursts (IT3) if the wavelength extends to decameter-kilometer. In this Letter, we investigate the probability of the flares that produce IT3, based on the sample of 2272 flares above M-class observed from 1996 to 2016. It is found that only 49.6\% of the flares are detected to be accompanied with IT3. The duration, peak flux, and fluence of the flares with and without IT3 both present power-law distributions in the frequency domain, but the corresponding spectral indices for the former (2.06$\pm$0.17, 2.04$\pm$0.18, and 1.55$\pm$0.09) are obviously smaller than that for the latter (2.82$\pm$0.22, 2.51$\pm$0.19, and 2.40$\pm$0.09), showing that the flares with IT3 have longer durations and higher peak fluxes. We further examine the relevance of coronal mass ejections (CMEs) to the two groups of flares. It is found that 58\% (655 of 1127) of the flares with IT3 but only 19\% (200 of 1078) of the flares without IT3 are associated with CMEs, and that the associated CMEs for the flares with IT3 are inclined to be wider and faster. This indicates that CMEs may also play a role in producing IT3, speculatively facilitating the escape of accelerated electrons from the low corona to the interplanetary space.
\end{abstract}


\section{Introduction} \label{sec:intro}
Solar flares are energetic phenomena in the solar atmosphere that can cause a sudden and rapid enhancement of electromagnetic radiation and efficiently accelerate electrons \citep{benz08}. In the CME/flare model \citep{1964NASSP..50..451C,1966Natur.211..695S,1974SoPh...34..323H,1976SoPh...50...85K,shibata95,lin04}, magnetic reconnection taking place in between the flare loops and the erupting CME is believed to play an important role in accelerating thermal electrons to a near-relativistic speed. After escaping from the reconnection site, these near-relativistic electrons primarily stream down to the dense lower atmosphere, giving rise to the hard X-ray emission at the footpoints of the flare loops through bremsstrahlung. At the same time, the plasma in the lower atmosphere is significantly heated and is then evaporated into the corona along the flare loops, causing the rapid enhancement of the soft X-ray emission \citep{fisher85}. 

On the other hand, a fraction of near-relativistic electrons move outward along open magnetic field lines and may produce solar type III radio bursts, which often appear as a bright and transient radio emission that quickly drifts from higher to lower frequencies with time \citep{2014RAA....14..773R}. The physical interpretation is that during the propagation of accelerated electrons, they interact with the ambient plasma and excite Langmuir waves, which then convert into radio emission \citep{Ginzburg1958}. As the frequency of induced emission is proportional to the square root of plasma density \citep[see][]{2002JGRA..107.1315C} that rapidly decreases as electrons move away from the Sun into the interplanetary space, the emission of type III bursts quickly drifts from hundreds of MHz to KHz in minutes. The counterpart of type III bursts with the frequency below tens of MHz is often named as interplanetary type III bursts (IT3). 

\begin{figure*}[h]
\gridline{\fig{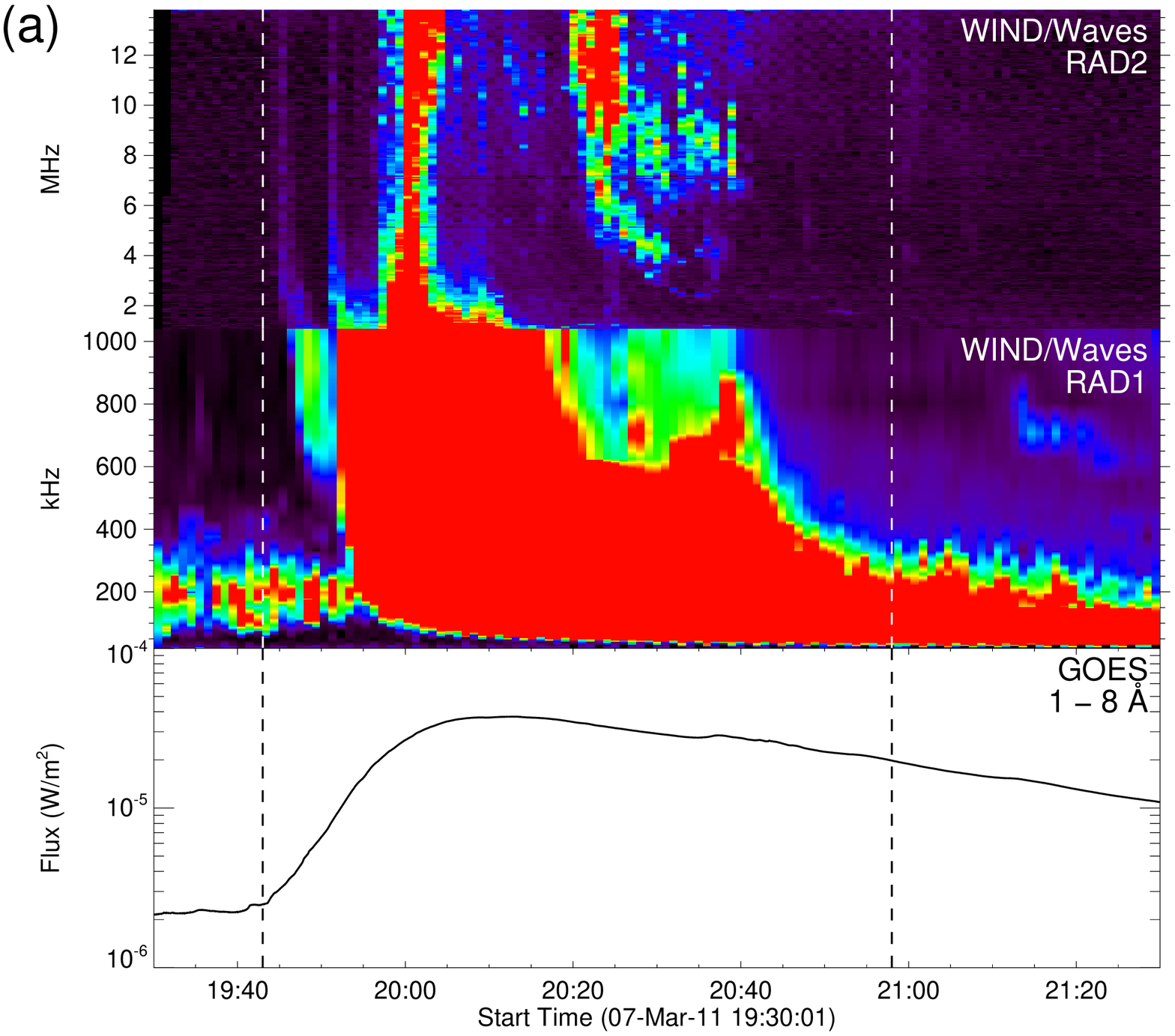}{0.48\textwidth}{}
		  \fig{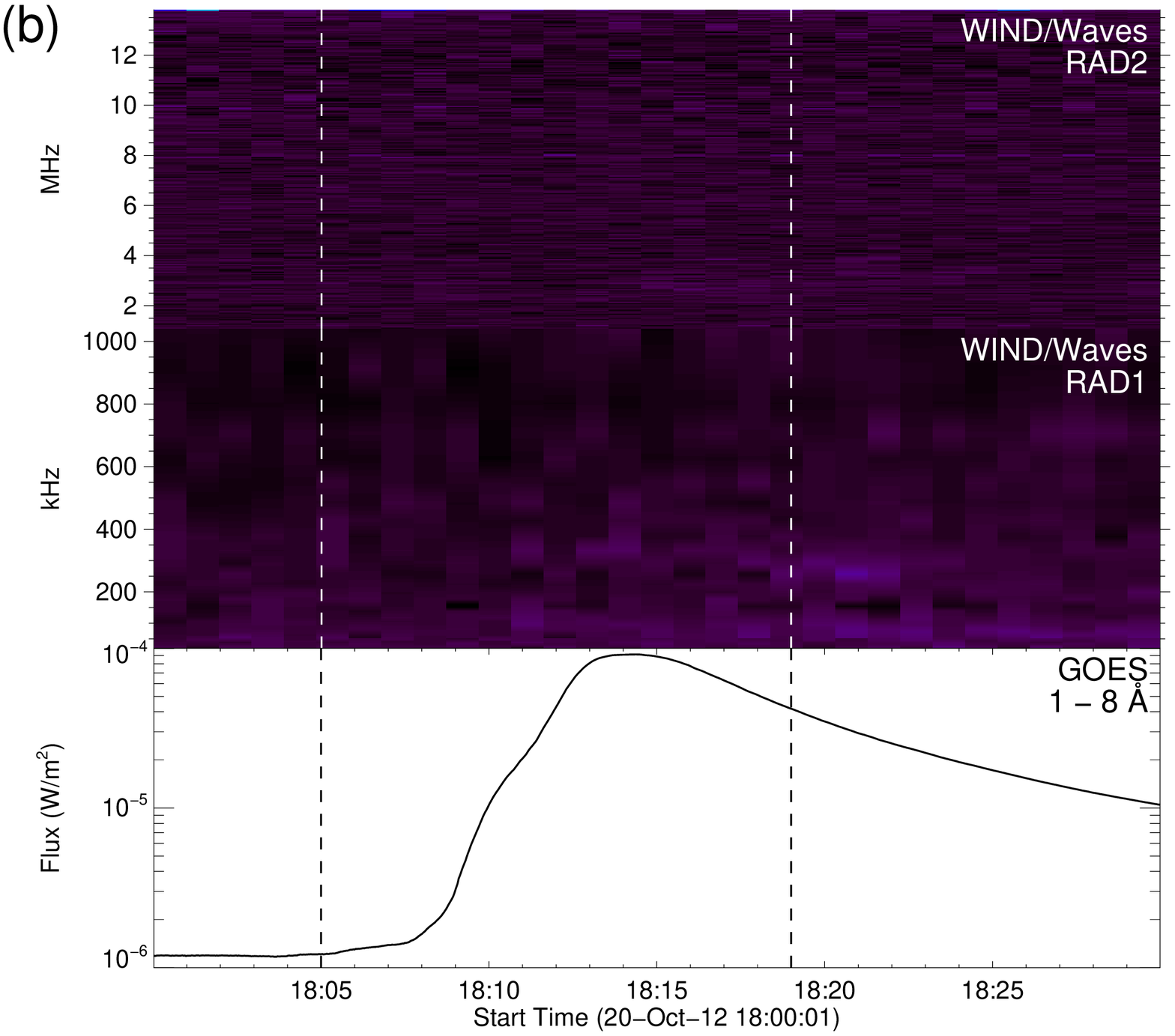}{0.48\textwidth}{}}\vspace{-0.03\textwidth}
\caption{(a) Dynamic spectrum from an interplanetary type III radio burst observed by WIND/WAVES and GOES soft X-ray 1--8 {\AA} flux of the associated flare observed on 2011 March 7. Two vertical dashed lines indicate the onset time (19:43 UT) and end time (20:58 UT) of the flare, respectively. The upper two panels clearly show the appearance of a number of type III bursts, which quickly drift from the frequency $>$ 10 MHz to $<$ 100 KHz. (b) The same as (a) but for the flare on 2012 October 20. In the latter case, one cannot find any signs of the associated type III bursts in the dynamic spectrum. \label{fig:exp_type3}}
\end{figure*}

The generation of IT3 requires an open flux, along which flare-accelerated electrons are able to propagate into the interplanetary space. This often takes place for flares occurring near the edge of active regions or coronal holes where the magnetic flux is primarily opened, frequently observed as jet-associated flares \citep[e.g.,][]{Krucker2011,Chennh2013,hong2017,chenbin2018,Glesener2018}. However, for most flares, especially for major ones ($>$M-class), they tend to appear near the center of active regions, where the background flux is mostly closed, thus preventing flare-accelerated electrons from escaping. Nevertheless, as suggested by \citet{Masson2013} and modeled by \citet{Masson2019}, the closed background flux can be partly opened if an interchange reconnection occurs between the closed flux and the ambient open flux such as coronal holes and streamers. This is highly possible for flares associated with the eruption of CMEs, as proved by many case studies \citep[e.g.,][]{vanDriel-Gesztelyi2008,Hillaris2011,cheng2018,zheng2017}. 

In spite of those case studies, there is still a lack of statistical study that can show how frequently the above scenario works in real events. To address this issue, in this Letter, we for the first time statistically investigate the relationship between major flares and IT3 taking place in the time period of 1996--2016. It is found that the frequency distributions of the flare parameters for the two groups of flares, i.e., with and without IT3, both present power laws but with different slopes. The more interesting thing is that the flares with IT3 have a much higher probability to be associated with CMEs and that the CMEs accompanied with flares and IT3 tend to be faster and wider. In Section \ref{sec:data}, we introduce the instruments we used. The data reduction and results are shown in Section \ref{sec:results}, which is followed by a summary and discussions in Section \ref{sec:summary}.

\begin{figure*}[h]
\plotone{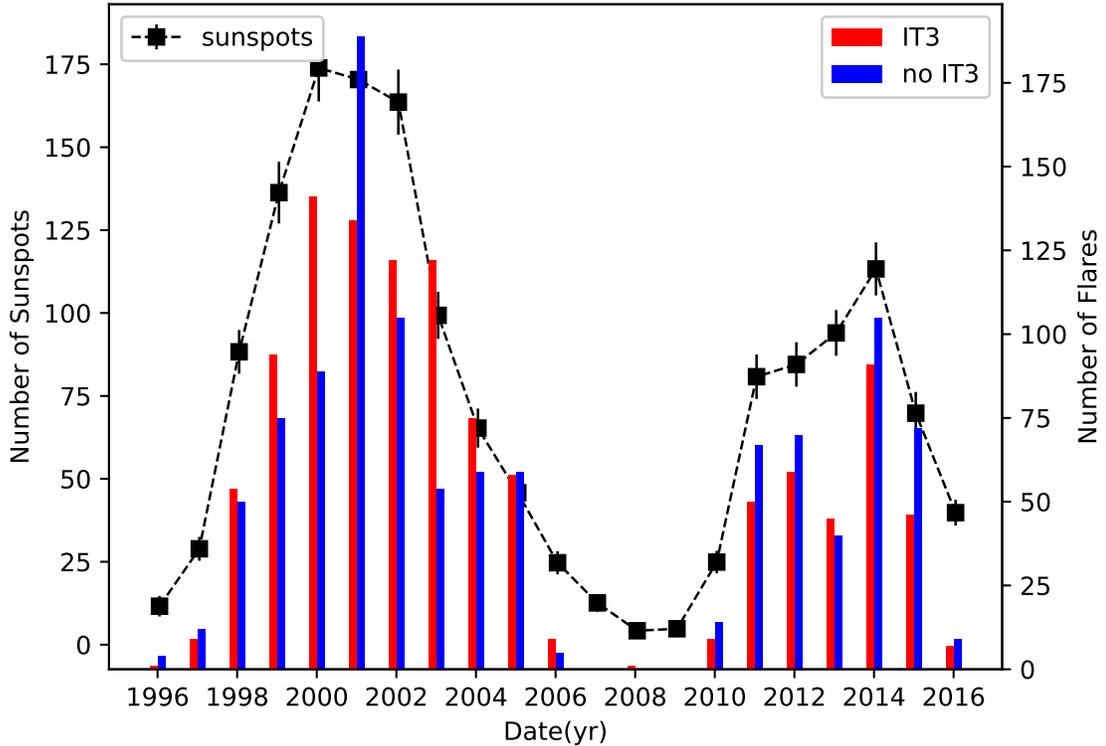}
\caption{Yearly variations of the numbers of sunspots (black square) and flares with (red) and without (blue) IT3 from 1996 to 2016. \label{fig:sunspot}}
\end{figure*}

\section{Instruments} \label{sec:data}

The flare list we used are based on the 1--8 {\AA} soft X-ray (SXR) flux data recorded by \emph{GOES} satellites. It contains basic information of each flare such as the onset, peak, and end times, heliographic location, as well as peak flux. Here, we only focus on major flares ($>$M-class) from 1996 to 2016 with a number of 2272 in total. To identify IT3, we use the dynamic spectrum data with a frequency range from 20 KHz to 14 MHz observed by \emph{WIND}/Waves \citep{Bougeret1995}. For most events, the IT3 are extensions of coronal type III bursts that start from a frequency larger than 14 MHz as shown in Figure \ref{fig:exp_type3}a. It indicates that the electrons emitting IT3 should be from the corona
($<$2.5 solar radius derived by the coronal density model of \citet{saito1977}), 
which we speculate originate from the same acceleration sites as that emitting hard X-rays \citep[e.g.,][]{Vilmer2002,Reid2014}. Moreover, as the flares we analyse are all above M-class, it is highly probable that the soft X-ray emissions are related to flare-accelerated energetic electrons, that is to say, the energetic electrons first heat the plasma in the flare loops, which then generate the soft X-ray emissions.

It is highly probable that the soft X-ray emissions are related to flare-accelerated energetic electrons, which partially heat the plasma in flare loops generating soft X-ray emissions, and partially escape into interplanetary space producing type III radio bursts.

For the sake of examining CMEs, white-light images provided by C2 and C3 of the Large Angle and Spectrometric Coronagraph \citep[LASCO;][]{brueckner95} on board \emph{SOHO} and COR1 and COR2 of the Sun Earth Connection Coronal and Heliospheric Investigation \citep[SECCHI;][]{howard08} on board \emph{Solar TErrestrial RElations Observatory} (\emph{STEREO}) are used. C2 and C3 (COR1 and COR2) have fields of view of 1.5--6 and 3.7--30 (1.1--4.6 and 4--15) solar radii, respectively. In order to associate flares and CMEs, we further take advantage of EUV images observed by the Extreme-Ultraviolet Imaging Telescope \citep[EIT;][]{del95} on board \emph{SOHO}, SECCHI/EUVI, and Atmospheric Imaging Assembly \citep[AIA;][]{lemen12} on board \emph{Solar Dynamics Observatory} (\emph{SDO}). The sunspot number of each year from 1996 to 2016 is obtained from WDC-SILSO, Royal Observatory of Belgium, Brussels.

\section{Data Reduction and Results} \label{sec:results}

\subsection{Correlation of flares to interplanetary type III bursts} \label{subsec:flare}

We first visually identify IT3 with WIND/Waves dynamic spectra for each major flare. The criteria for defining a flare that is associated with an interplanetary type III burst are as follows: there appear one or more type III bursts whose onset times are in the time interval of the flare (from onset to end), and the corresponding frequency of the bursts extends to below 400 KHz. Two examples are shown in Figure \ref{fig:exp_type3}. For the flare took place on 2011 March 7 (Figure \ref{fig:exp_type3}a), one can see that, as the flare started, a type III burst group consisting of many reversed J-shape structures shortly appeared in the dynamic spectrum, showing a rapid drift of the radio emission toward lower frequencies. Each reversed J-shape structure corresponds to one type III burst caused by a beam of electrons.  At the flare end time, all type III bursts have merged together at the frequency of $\sim$ 0.2 MHz, corresponding to the plasma number density of 500 cm$^{-3}$. In contrast, for the flare on 2012 October 20 (Figure \ref{fig:exp_type3}b), we cannot find any signs of type III bursts in the dynamic spectrum in the frequency range of 0.1--14 MHz during the flare. 
 
For most flares, identifying their association with IT3 is not difficult. However, for a small number of flares, we cannot make a decision with a high confidence, which is mainly limited by the fact that some IT3 may appear during two successive flares. Of course, like any other studies, we also cannot get rid of the possibility of an interplanetary type III burst is too weak to be detectable. In terms of their association with IT3, all flares are divided into two groups: one of 1127 flares (49.6\% of 2272) with and the other of 1078 flares (47.5\% of 2272) without IT3. The remaining 67 (2.9\% of 2272) flares cannot be classified due to the lack of WIND/Waves data. A text file that shows our classification for the whole sample is available from the website\footnote{https://astronomy.nju.edu.cn/szll/szgk/fjs/20191102/i44994.html}, in which the reliability of identification is marked with a number between 3 and 0, namely, 3 for flares definitely with, 2 for flares likely with, 1 for flares likely without, and 0 for flares definitely without IT3. Thus, the flares marked with 3 and 2 belong to the first group and those with 0 and 1 to the second group.

The numbers of two groups of flares as a function of time are displayed in Figure \ref{fig:sunspot} together with the sunspot number. One can find that they are closely related to each other. Near the solar minimum, the sunspots almost disappear, and the occurrence rate for both types of flares is very low. Only a few number of events take place in the minimum year. When the solar cycle proceeds, the numbers of two types of flares start to increase, reach a peak, and then decrease, almost synchronously with the sunspot number. Moreover, it is noticed that the number of the first group of flares (with IT3) is slightly larger than that of the second group (without IT3) in most years. However, in each year of the 24th cycle except for 2013, the number of the first group is moderately smaller than that of the second group. Overall, the occurrence rate for two types of flares in the time period under study is very similar.

\begin{figure}
\gridline{\fig{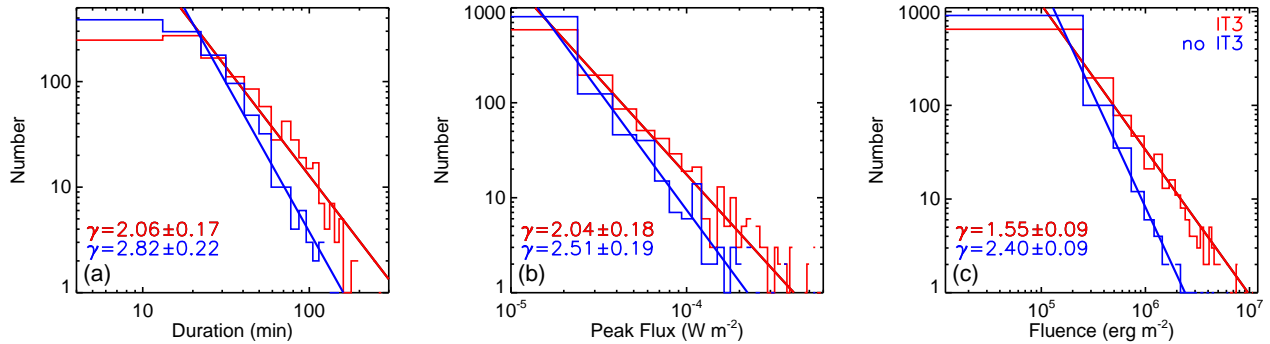}{1\textwidth}{}} \vspace{-0.04\textwidth}
\caption{Frequency distributions of the flare duration (a), peak flux (b), and fluence (c) for the flares with (red) and without (blue) IT3. The oblique lines in each panel represent the power law fits with the spectral indices indicated at the lower left corner.}\label{fig:distribution1}
\end{figure}

\begin{figure*}[h]
\gridline{\fig{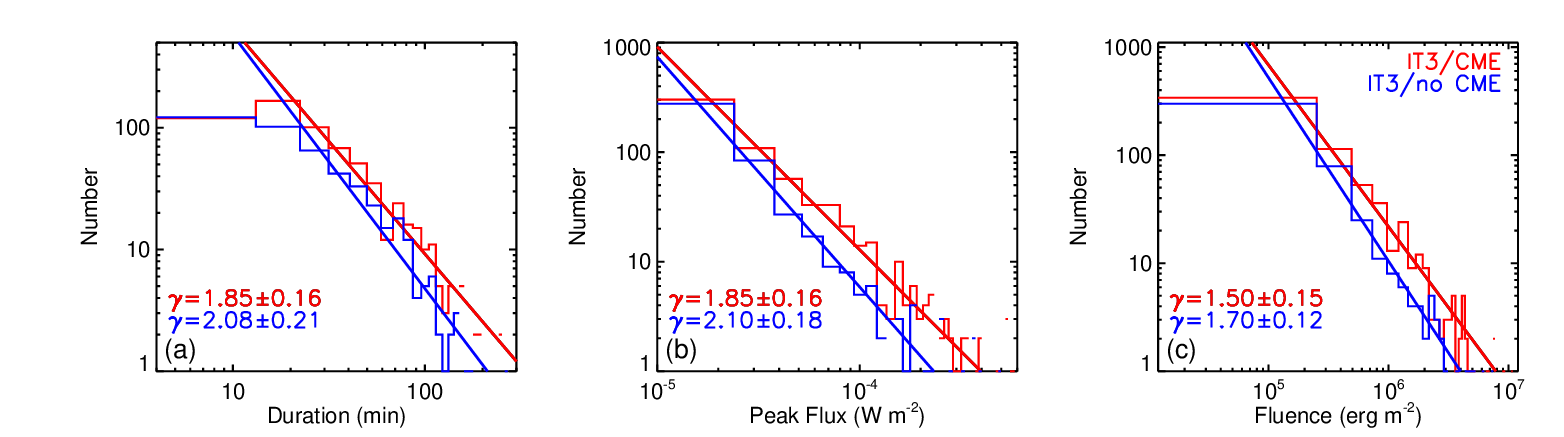}{1\textwidth}{}}\vspace{-0.06\textwidth}
\gridline{\fig{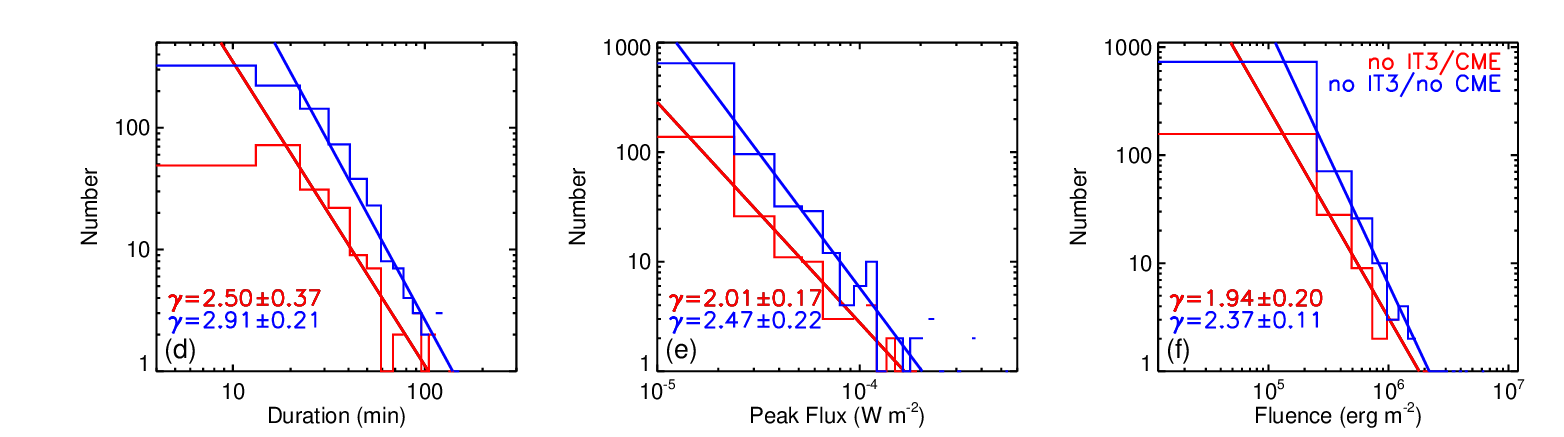}{1\textwidth}{}}\vspace{-0.04\textwidth}
\caption{Frequency distributions of the flare duration (a), peak flux (b), and fluence (c) for IT3-associated flares with (red) and without (blue) CMEs. (c)--(f) are the same as (a)--(c) but for IT3-unassociated flares. The oblique lines in each panel represent the power law fits with the spectral indices shown at the lower left corner.}\label{fig:distribution2}
\end{figure*}

To differentiate the two groups of flares, we further examine the frequency distributions of the flare parameters including the duration, peak flux, and fluence (total flux). Based on the GOES SXR 1--8 {\AA} flux of the flares, the duration is computed by the end time minus onset time, the peak flux is obtained directly from the flare list, and the fluence is simply estimated by multiplying the duration with half of the peak flux. The results are shown in Figure \ref{fig:distribution1}. One can clearly see that the occurrence ratio of the flares with IT3 to the flares without IT3 significantly increases with the flare duration, peak flux, and fluence. For a relatively long duration ($>$100 mins) and/or high peak flux ($>2\times10^{-4}$\ W/m$^2$), almost all flares are associated with IT3. Moreover, it is found that the frequency distributions of the duration, peak flux, and fluence all present power laws. However, the interesting thing is that the corresponding spectral indices are obviously distinct in the two groups; they are 2.06$\pm$0.17, 2.04$\pm$0.18, and 1.55$\pm$0.09 for the flares with IT3, markedly smaller than 2.82$\pm$0.22, 2.51$\pm$0.19, and 2.40$\pm$0.09 for the flares without IT3. The spectral indices for the flares with IT3 are also found to be smaller than the values (2.93$\pm$0.12, 2.11$\pm$0.13, and 2.03$\pm$0.09) in \citet{veronig2002} and those (2.87$\pm$0.09, 2.16$\pm$0.03, and 2.01$\pm$0.03) in \citet{yashiro2006} for all flares. The results show that the flares with IT3 are a subset that consists of preferentially larger events. In other words, a flare with a longer duration, higher peak flux, and larger fluence is more likely to be associated with an IT3. 

It is worth mentioning that the spectral indices for the peak flux and fluence could be overestimated as no pre-event background is subtracted \citep{Aschwanden2011}. However, the lack of the background subtraction will not influence our main conclusions because of two reasons: (1) the background for major flares should be trivial \citep[e.g.,][]{sadykov2019} and (2) we focus more on the comparison between different groups of flares rather than specific subsets.

\begin{table}
\caption{Spectral indices of flare parameters for four groups of flares.}
\label{tb1}{
\hspace{0.2\textwidth}
\scalebox{1}[1]{
\begin{tabular}{cccc}
\\ \tableline \tableline
Groups             & Duration        & Peak Flux               &Fluence  \\
\hline
IT3/CME           &1.85$\pm$0.16  & 1.85$\pm$0.16           & 1.50$\pm$0.15 \\
IT3/no CME      &2.08$\pm$0.21  & 2.10$\pm$0.18           & 1.70$\pm$0.12 \\
no IT3/CME      &2.50$\pm$0.37  & 2.01$\pm$0.17           & 1.94$\pm$0.20 \\
no IT3/no CME &2.91$\pm$0.21  &2.47$\pm$0.22           & 2.37$\pm$0.11 \\
\tableline
\end{tabular}}}\\
\end{table}

\subsection{Correlation of CMEs to interplanetary type III bursts} \label{subsec:cme}
In this Section, we study the role of CMEs in the process(es) of flares producing IT3. The appearance of CMEs for each flare is carefully inspected using the CDAW CME catalog\footnote{https://cdaw.gsfc.nasa.gov/}. For the time interval of 1996--2006, \citet{Cheng_2010} have checked the association between flares and CMEs. Their method is to visually examine the white-light images of the LASCO at a fixed time window of 2 hours after the flare peak and then further identify the locations of the source regions with transient brightenings, large-scale dimmings, and waves. The same procedure is applied to the events in the time range of 2007--2016. An improvement is that we also take advantage of the SECCHI and AIA data, in comparison to \citet{Cheng_2010}, who only used the LASCO and EIT data. Thanks to the higher cadence and spatial resolution of the modern data, especially the AIA data, the degree of accuracy to associate flares and CMEs is improved significantly. In addition, to identify CMEs more reliably, we also check whether there is an erupting structure that continuously evolves from the lower to higher corona, for example, from the field of view of the AIA to that of the LASCO. 

We find that, from 1996 to 2006, 55\% (453 of 819) of the flares with IT3 are associated with CMEs; while for the flares without IT3, only 20\% (141 of 701) are CME-associated events, and the remaining majority are CME-less events. From 2007 to 2016, among 307 flares with IT3, 66\% (202 of 307) are associated with CMEs; by comparison, among 377 flares without IT3, only 16\% (59 of 377) are associated with CMEs. To combine all the events from 1996 to 2016, 58\% of IT3-associated flares are accompanied with CMEs, 41\% are CME-less events. The proportion of CME association is much higher than the corresponding proportion of 19\% for the flares without IT3. It shows that the flares with IT3 are more likely to be accompanied by CMEs.

In terms of correlation to CMEs, we further divide all flares into four groups: the first group of flares are associated with both IT3 and CMEs, the second one with IT3 but no CMEs, the third one with CMEs but no IT3, and the fourth one without IT3 and CMEs. The frequency distributions of the flare duration, peak flux, and fluence for the four groups are shown in Figure \ref{fig:distribution2}. It is obvious that the three parameters for each group are still power-law distributed. Moreover, we also find that, among the four groups of flares, the spectral indices are the smallest for the group with both IT3 and CMEs but the largest for the group without IT3 and CMEs (Table \ref{tb1}). This is comparable with \citet{yashiro2006} who found that the spectral indices are smaller for flares with CMEs than for ones without CMEs. The spectral indices of the flare duration, peak flux, and fluence derived here are 1.85$\pm$0.16, 1.85$\pm$0.16, and 1.50$\pm$0.15, respectively, which are obviously smaller than the values (2.49$\pm$0.11, 1.98$\pm$0.05, and 1.79$\pm$0.05) for the flares only with CMEs derived by \citet{yashiro2006}. 

\begin{figure*}
\gridline{\fig{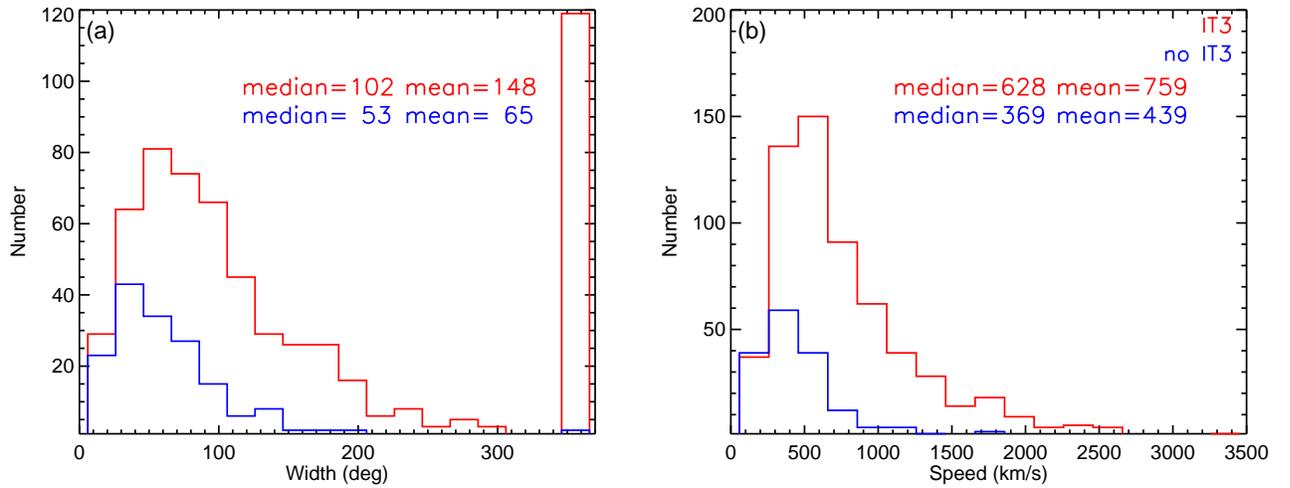}{1.0\textwidth}{}}\vspace{-0.05\textwidth}
\caption{Distributions of the width (a) and velocity (b) of associated CMEs for the flares with (red) and without (blue) IT3.}\label{fig:distribution3}
\end{figure*}

We also compare the distributions of the apparent width and velocity of associated CMEs for the first and third groups of flares as shown in Figure \ref{fig:distribution3}. It is evident that for the flares that produce IT3, the associated CMEs tend to be wider and faster. The apparent width mainly varies in the range of 40$^{\circ}$--360$^{\circ}$ with the average and median values of 148$^{\circ}$ and 102$^{\circ}$, respectively. If the halo CMEs, whose width equals to 360$^{\circ}$, are excluded, the average and median values become to be 97$^{\circ}$ and 84$^{\circ}$, respectively. The apparent velocity is primarily within a range of 400--2500 km s$^{-1}$ with the average and median values of 759 and 628 km s$^{-1}$, respectively. These values are significantly greater than the widths (65$^{\circ}$ and 53$^{\circ}$) and the velocities (439 and 369 km s$^{-1}$) for the flares without IT3. This suggests that CMEs may play a role in, at least be relevant to, the process(es) of the flares generating IT3. 

\section{Summary and Discussion} \label{sec:summary}
In this paper, we make a statistical study on the relationship between flares and IT3 and find that about half of major flares from 1996 to 2016 are accompanied with IT3. The number of the flares either with or without IT3 varies with the solar cycle and is not prominently distinct to each other in each year. We also study the frequency distributions of the flare duration, peak flux, and fluence and find that they all present power laws. The spectral indices for the flares with IT3 are derived to be 2.06$\pm$0.17 for the duration, 2.04$\pm$0.18 for the peak flux, and 1.55$\pm$0.09 for the fluence, respectively, which are obviously smaller than the values of 2.82$\pm$0.22, 2.51$\pm$0.19, and 2.40$\pm$0.09 for the flares without IT3. It shows that the flares producing IT3 tend to last for a longer time and to be more powerful. This supports the conclusion of \citet{Reid2017} that the flares with higher peak flux of hard X-rays ($>$25 keV) was strongly correlated with the occurrence of IT3s. In principle, this is expected because the high-speed electrons that stream down to produce the X-ray emission of flares are generally related to the electrons that move upward to form IT3. It is even documented that these two beams of electrons are most likely from the same acceleration regions 
\citep[e.g.,][]{Vilmer2002,Reid2014}. Thus, the stronger a flare is, the higher its IT3 occurrence rate is.

Energetic electrons and open flux are two necessary conditions for the production of IT3. In the standard flare/CME model, electrons are believed to be mainly accelerated at the reconnection region located in between the flare loops and the erupting CME. For the upward electrons, they should be first injected into the CME flux rope, which is a closed flux and thus prevents electrons from escaping. However, such a constraint of the CME flux may be invalid if an interchange reconnection takes place between the erupting CME flux and the ambient open flux, as suggested by \citet{Masson2013,Masson2019} and supported by many case studies \citep[e.g.,][]{vanDriel-Gesztelyi2008,Hillaris2011,cheng2018,zheng2017}. 

We find that 58\% of the flares with IT3 are also associated with CMEs, and that the proportion is only 19\% for the flares without IT3. For the flares that are associated with both IT3 and CMEs, the spectral indices of the duration, peak flux, and fluence even decrease to 1.85$\pm$0.16, 1.85$\pm$0.16, and 1.50$\pm$0.15, respectively, which are the smallest among all groups of flares. Moreover, the CMEs associated with both flares and IT3 are evidently wider and faster than those associated with flares only. This implies that powerful CMEs tend to have a higher probability of taking place reconnection between the CME flux and the ambient open flux so as to release flare-accelerated electrons. The statistical result presented here is in favor of the scenario of interchange reconnection, which may occur more frequently than previously thought. 

For the flares producing IT3 but without CMEs, it is suspected that their source regions are already close to the open flux so that no further interchange reconnection is required. In this case, accelerated electrons can escape easily to the interplanetary space along the open flux, producing the IT3 \citep[e.g.,][]{li2009,Krucker2011,chenbin2018,Glesener2018}. Statistically, the former scenario seems to occur slightly more frequently than the latter one, as evidenced by the proportion of 58\% for CME-associated flares vs. 41\% for CME-less flares that are accompanied by IT3. 

For the flares without IT3, open flux that facilitates the formation of IT3 is less likely to be formed as suggested by the low occurrence rate (19\%) of CMEs. In fact, the few CMEs in such cases are narrow and slow, which are relatively unfavourable for initiating interchange reconnection.
It is also possible that no enough energetic electrons escape to the interplanetary space to produce IT3 or the induced radio emission is too weak to be detected.

\acknowledgments We acknowledge helpful suggestions by the referee, which significantly improve the manuscript. We used the IDL programs provided by WIND/WAVES to make dynamic spectrum of type III radio bursts. This work is supported by NSFC grants 11722325, 11733003, 11790303, 11790300, 11673012, Jiangsu NSF grant BK20170011, the ``Dengfeng B" program and College Students' Innovative Training Program of Nanjing University, and the Alexander von Humboldt foundation. 


\end{document}